\documentclass[12pt,preprint]{aastex}
\begin{document}

\title{Stellar Bowshocks in the Northern Arm of the Galactic Center:
More Members and Kinematics of the Massive Star Population}

\author{A. Tanner, A. M. Ghez, and M. R. Morris}
\affil{UCLA Department of Physics and Astronomy, Los Angeles, CA 90095-1562}
\author{J. C. Christou}
\affil{Center for Adaptive Optics, Santa Cruz, CA}
\authoremail{tanner@astro.ucla.edu}

\begin{abstract}
We present new 2.2 $\micron$ diffraction-limited images 
from the W. M. Keck 10 m and Gemini 8 m telescopes of the cool
Galactic Center sources, IRS 1W, 5, 8, 10W, and 21 along with new 
proper motions for IRS 1W, 10W and 21. These observations were carried out
to test the bowshock hypothesis presented by Tanner et al. as an alternative to a very recent (10$^4$ yr) epoch of
star formation within the tidal stream of gas and dust known as the Northern Arm. 
Resolved asymmetric structure is detected in all the sources, with 
bowshock morphologies associated with IRS 1W, 5, 8 and 10W. For IRS 
1W and 10W, there is an agreement between the
position angle of the asymmetry and that of the relative velocity vector of the
near-infrared source with respect to the Northern Arm gas strengthening
the bowshock hypothesis. We therefore 
conclude that the observed morphology is indeed a bowshock generated by sources plowing through the
Northern Arm. 
Furthermore, the large extent of the resolved structures (310-1340 AU) along with their 
luminosities ($\sim$10$^{4-5}$ L$_{\sun}$) 
suggests that their central sources are Wolf-Rayet stars,
comparable to the broad He emission-line stars, which have strong winds on the order of 1000 km s$^{-1}$. 
The bowshock morphology, along with the proper motion measurements, provide three-dimensional
orbital solutions for this enigmatic class of objects; IRS 1W and 10W have 
orbital planes that are consistent with that of the putative clockwise 
plane which has been proposed as a solution for the He I emission-line stars.
While these observations eliminate the need to invoke star formation within the Northern Arm, they 
increase by 14\% the total known population of
massive, young stars with strong winds, whose origin remains unexplained in 
the context of the nearby supermassive black hole. 
\end{abstract}

\altaffiltext{1}{Also affiliated with UCLA IGPP}

\section{Introduction}

Within the Galaxy's central cluster there are
a number of enigmatic sources (IRS 1W, 2, 5, 8, 10W and 21)
that have eluded classification for almost three decades (Rieke \& Low 1973, Becklin and Neugebauer 1975).
They are all spatially coincident with the Northern Arm, a tidal stream of dust and 
gas that has been mapped at both near-infrared and radio wavelengths and appears to be 
infalling and/or orbiting around the Galaxy's central supermassive black hole
(Lo \& Claussen 1980; Serabyn \& Lacy 1985; Lacy et al. 1979; Herbst et al. 1993; 
Paumard, Maillard \& Morris 2004).  These sources also all have
spectral energy distributions that peak near 10 $\micron$, high luminosities 
(L$\sim$10$^5$ L$_{\sun}$, Becklin \& Neugebauer 1978), and, in the    
cases of IRS 1W and 21, nearly featureless near-infrared spectra (Moultaka et al. 2004; Krabbe et al. 1995)
and polarization properties that cannot be accounted for by the intervening ISM
(Eckart et al. 1995; Ott et al. 1999).  Based on these observed properties,
many have suggested that these Northern Arm sources are 
very recently formed stars that are being transported into the central core 
along with the Northern Arm inflow (Krabbe et al. 1995; Blum et al. 1996; Ott et al. 1999).
As pointed out by Tanner et al. (2002), this scenario is problematic since the short Northern Arm timescales
require that the stars would have to have formed, 
or to have begun forming, prior to the infall of the Northern Arm.
In this case one would not expect all of the stars to still be embedded in the gas of the 
Northern Arm, given that the newborn stars will immediately 
follow ballistic orbits, while the gas in which 
they formed will be subject to strong, additional, non-gravitational forces. 

We recently argued, based on a simple radiative transfer model of the spatially resolved 2-25 $\micron$
images and photometry of IRS 21, that all the Northern Arm sources
are massive stars experiencing rapid mass loss, as either Wolf-Rayet (WR) or asymptotic giant branch (AGB) stars (Tanner et al. 2002). 
As the Northern Arm flows past these stars,
the interaction between the material in the Northern Arm and the intrinsic stellar winds
generates a bowshock which produces the observed spatially resolved 
emission (Ott et al. 1999; Tanner et al. 2002).
While the resolved structure around IRS 21 lacks the characteristic horseshoe 
morphology, this could simply be a projection effect, if the relative
velocity between the outflow source and the Northern Arm is oriented near the
line of sight. One appeal of the bowshock model is that it explains the
association of these sources and the Northern Arm without requiring that the stars 
form within the Northern Arm before or during the infall toward
the central black hole. A bowshock model instead allows for the much more
plausible circumstance that the sources are simply in the path of the infalling 
Northern Arm gas and dust, allowing them to be members of the observed population of
stars seen in and around the central parsec. This hypothesis is supported by recent
high-angular-resolution images of IRS 8, which show a clear bowshock 
morphology around this star (Geballe et al. 2004).

In this paper, we present new near-infrared observations for
the remaining Northern Arm sources, IRS 1W, 5, and 10W, as well as a new analysis of IRS 21 and 8
in order to test the bowshock hypothesis.  Section 2 describes
the data sets that were obtained for this work, Section 3 reviews the techniques used in the image construction and
deconvolution, and
Section 4 presents the measured morphologies and proper motions for the 
surveyed sources. Section 5 makes the case that these sources are Wolf-Rayet stars -- young, massive, post-main sequence stars -- generating bowshocks 
as they plunge through the Northern Arm, and it discusses the implications of their source classification. 

\section{Observations}

\subsection{Speckle Imaging with Keck I/NIRC \label{nircobs}}

Speckle imaging observations of IRS 1W, 5, 10W, and 21 
were obtained in the
K bandpass ($\lambda_o$=2.2 $\micron$, $\Delta\lambda$=0.43$\micron$) 
 using the W. M. Keck I 10-meter telescope and the facility near-infrared camera (NIRC; Matthews \& Soifer 1994; Matthews et al. 1996).
The camera has a pixel scale of 20.400$\pm$0.042 mas/pixel (Ghez et al. 2004) and a 
correspondingly small field of view (FOV) of 5$''\times$5$''$, limiting observations to one target at a time. 
Each object was observed in sets of $\sim$200$\times$0.14 second exposures. 
The long-exposure seeing averaged 0$\farcs$6 during these observations.

All sources except IRS 5 were observed on multiple nights (see Table~\ref{obs}) as part of a proper
motion campaign (Ghez et al. 1998, 2000, 2004). 
Measurements of the 5''$\times$5'' field centered on Sgr A* began in 1995, providing a eight-year time
baseline for measurements of the proper motion of IRS 21. Beginning in 1998, additional images 
were obtained to mosaic the central stellar cluster frames together with those that 
contained the SiO masers, IRS 7 and 10EE, in order to register the radio and
infrared reference frames (see Ghez et al. 2004), resulting in a four-year time baseline for IRS 1W and 10W. 

\subsection{AO Imaging with Gemini/Hokupa'a+Quirc$^1$\label{gemobs}}

We have also utilized public Gemini Adaptive Optics (AO) 2.2 $\micron$ images$^1$ of IRS 8 collected on 
2000 July 02 using the Hokupa'a AO system and the QUIRC near-infrared camera with a FOV of 20''$\times$20'' and a plate 
scale of 19.745$\pm$0.090 mas/pixel. The plate scale estimate for these images is based on
the average and standard deviation of the independently determined plate scales estimated using the NIRC 
plate scale and the five offsets from IRS 16C to IRS 16NW, 16SW, 16NE, 
33N, and IRR1, compared to those predicted by Ghez et al. (2004). 
In these observations, USNO 0600-28577051, which has an R (0.7$\micron$)-band magnitude of 13.8 and is 19$\arcsec$ away from 
Sgr A*, served as the natural guide star.

\footnotetext[1]{Based on observations obtained at the Gemini Observatory, which is
operated by the Association of Universities for Research in Astronomy,
Inc., under a cooperative agreement with the NSF on behalf of the
Gemini partnership: the National Science Foundation (United States),
the Particle Physics and Astronomy Research Council (United Kingdom),
the National Research Council (Canada), CONICYT (Chile), the
Australian Research Council (Australia), CNPq (Brazil) and CONICET
(Argentina).}

\section{Data Analysis} 

\subsection{Image Construction}

The final images for all the observations are produced with the same
basic two steps:  1) each frame is sky-subtracted, flat-fielded,
bad-pixel-corrected, and resampled by a factor of two (except for the Gemini data), and 2) the 
images at a given telescope pointing are combined by shift-and-adding (SAA) on the brightest source
located near the center of the field of view. For the speckle images,
the centroid of the dominant speckle within the speckle interferometry pattern is chosen and for the adaptive optics
images, the centroid of the star over a region corresponding to the diffraction
limit was used. These SAA maps have Strehl ratios of $\sim$2-5\% and 15\% for the
Keck speckle images and the Gemini AO images,
respectively, and it is these images that are deconvolved (see Section 3.2) to study the morphology
of the Northern Arm sources.

For the proper motion analysis, the Keck speckle SAA maps at different
telescope pointings are assembled using all the bright sources in the overlap
regions between pointings (typically 20-30 sources that are
brighter than K$\sim$14 mag).  The final mosaics cover roughly
10''$\times$10''.  Source positions, both final positions and those used to register the
different pointings, are based on peaks in the cross-correlation of the
SAA maps with the diffraction-limited PSF core as described by
Ghez et al. (1998); here, a cross-correlation threshhold of 0.7 is used.
The positions of all the identified sources within the mosaiced images covering eight separate epochs are aligned with respect to 
a reference image (1999 July) by minimizing the net offset of all the stars within the mosaic. 
The NIRC data set provides 21, 10, and 7 independent astrometric position measurements for IRS 21, 1W, and 10W, respectively. 

\subsection{Deconvolution}

In order to enhance our sensitivity to the detailed structure of any extended emission, 
we removed the effects of the residual seeing halo from both the speckle SAA and the
adaptive optics images by applying a deconvolution algorithm to selected regions containing each of the Northern Arm sources.
After investigating a variety of deconvolution routines, including blind 
deconvolution (IDAC, Jefferies \& Christou 1992; Christou et al. 1999), 
Lucy-Richardson (Lucy 1974; Richardson 1972), maximum-entropy and 
maximum-likelihood algorithms (Press et al. 1983),
we concluded that the IDAC deconvolution routine provides the most robust results, 
based on the consistency of the observed extended morphology between 
maps taken in different years. This physically constrained
algorithm\footnote{The current version of the code, implemented by E.K. Hege and
M. Chesalka, is available from the Center for Adaptive Optics webpages at http://cfao.ucolick.org/software/idac/} 
was initially described by Jefferies \& Christou (1992) and its application to AO
imaging by Christou et al. (1999).
In summary, IDAC is a multi-frame blind deconvolution package which makes use of multiple observations
of the same source, where the differences between the observations are assumed to be due solely to differences in the
point spread function (PSF), in order to solve for the underlying object distribution function as well as the PSF. 
The solution is driven by
minimizing, using conjugate gradient techniques, an error
metric that characterizes the difference between the observations and the convolution of the target and PSF
estimates (See Appendix A for more details.). 

\section{Results}

\subsection{Morphology \label{morph}}

Figure~\ref{nearall} shows contour plots of the 0$\farcs$5$\times$0$\farcs$5 region for the 
sources included in this study, all of which are clearly extended when compared to a neighboring PSF.
IRS 1W, 5, 8, and 10W show signs of the horseshoe morphology expected from a bowshock. 
IRS 21 shows an azimuthally symmetric extension, which is consistent with both the morphology and interpretation of a face-on
bowshock as was deduced from the non-deconvolved images in Tanner et al. (2002). The bowshock hypothesis is, therefore, well
supported based on a simple qualitative analysis of the images.

To estimate the extent of the bowshock structures, the central compact 
component is subtracted from each deconvolved image (see Appendix and Figures~\ref{morphcalc} and~\ref{allbow}).  An initial estimate of the 
angular extent of the diffuse emission, r$_i$, is then determined from 
the angular distance between the centroid position of the central source 
(prior to subtraction) and the position of the peak in the diffuse emission 
(after subtraction).  As illustrated in Figure~\ref{morphcalc}, this 
distance is used to quantify the two-dimensional structure by extracting 
radial profiles averaged over an azimuthal angle, 
$\theta_{avg}$(rad)=0.05/r$_i$(arcsec).
This $\theta_{avg}$ corresponds to 25, 25, 10 and 25 degrees for 
IRS 1W, 5, 8 and 10W, respectively.
For those radial profiles along which the peak is $>$30\% of the overall 
maximum of the diffuse emission, the radius at which each profile reaches 
its peak intensity is recorded and its corresponding uncertainty is 
estimated by fitting a gaussian to the radial profile (see Table~\ref{radii}).

The extent of these bowshocks allows us to apply models of the shape of 
the shock front to the observed morphologies.  If the bowshocks are being 
produced from the flow of dust and gas along the Northern Arm encountering 
the wind from the central star, then a plane-parallel flow incident on a 
spherical wind would be the appropriate model.  While the dust density of 
the Northern Arm is quite complex over large scales, we assume that over 
small scales it can be modeled as a constant density medium. 
In this case, the shape of a bowshock can be described analytically as 
$R_{3-D}(\hat{\theta})=R_{o}csc\theta\sqrt{3(1-\theta cot\theta})$, where 
$R_o$ is the stand-off distance and 
$\theta$ = $cos^{-1}\hat{\theta}\cdot\hat{\theta_o}$ is the angle between 
any direction $\hat{\theta}$ (unit vector) and the unit vector 
$\hat{\theta_o}$ describing the direction of the relative velocity between 
the central star and the incident medium, $\vec{v}_{rel}$
(Wilkin 1996; Canto et al. 1996).  In order to fit our two-dimensional 
data set to this three-dimensional model, we consider only the points 
that lie in the plane defined by 
$\vec{v}_{rel}$ and $\vec{v}_{rel}\times\vec{s}$, where $\vec{s}$ is the 
line of sight.  In this plane, $\vec{R}_{2-D}=(R_{3-D\vec{v}_{rel}}\hat{\theta_o},R_{3-D\vec{v}_{rel}\times\vec{s}}\hat{i}_{\vec{v}_{rel}\times\vec{s}})$, 
where the $\hat{i}$ is a unit vector in the direction given by the
subscript.  Next, we project this two-dimensional model onto the plane of 
the sky, according to the inclination angle, i, the angle between 
$\vec{v_{rel}}$ and the line-of-sight.  This gives 
$\vec{R}_{2-D, sky}=(R_{3-D \vec{v}_{rel}}\hat{\theta}_{o,sky}sini,R_{3-D\vec{v}_{rel}\times\vec{s}}\hat{i}_{\vec{v}_{rel}\times\vec{s}})$, 
where $\hat{\theta}_{o,sky}$ is $\hat{\theta}_o$ projected
onto the plane of the sky 
\footnote{ This procedure projects onto the line of sight only the intersection of 
the bowshock with a plane passing through the central star, the apex 
of the bowshock and the vector $\hat{s}$.  It is thus only an approximation to
the projected brightness distribution, a better approximation for which
would be the curve represented by those lines of sight which are
tangent to the 3-dimensional bow shock surface.}. We use $\chi^2$ minimization to 
find the best fitting bowshock model to the data presented in 
Table~\ref{radii}.  This provides estimates of the standoff distances, 
R$_o$, and the projected orientation of $\hat{\theta_o}$, the latter of 
which is described by the position angle on the plane of the sky,  
PA=tan$^{-1}$(R$_{\alpha}(\hat{\theta}=\hat{\theta_o})$/R$_{\delta}(\hat{\theta}=\hat{\theta_o})$) and the inclination. 

Figure~\ref{allbow} shows the best fitting bowshock models for IRS 1W, 5, 8, and 
10W  while Table~\ref{bowprop} gives the best-fitting parameters and their 
uncertainties, which are estimated from the change in the value of the parameter 
which results in a variation of the $\chi^2$ value by 1.  When we try to fit this simple
bowshock model to IRS 1W, it was discovered that a better fit to the data could be achieved if the
position of the apex of the model is displaced 200 AU to the North. While this would reduce the
value of the standoff distance by the same amount, the value of the inclination of the best fitting
model remains 90 degrees. We plot both model fits in Figure~\ref{irs1wcomp} for comparison and
use a value of 90 degrees for the inclination of the bowshock in subsequent calculations. The shift
in the apex is of no concern as it could be explained as an offset in the position of
the photocenter due to winds from neighboring sources while the poor fit to the data in the wings of the
bowshock for the unshifted apex is due to the simplicity of the bowshock model which does
not account for a density gradient or the magnetic field within the Northern Arm. 

\subsection{Proper Motion of the Northern Arm Sources}

In order to investigate the relative motion occurring between the embedded sources and the Northern Arm, 
the proper motions of IRS 21, 10W, and 1W  
are measured from the multi-epoch 2.2 $\micron$ speckle data set. 
The proper motions are estimated from a linear least-squares fit through their positions
as a function of time. The uncertainties on the positions are estimated by assuming that
the errors are equal for each source and in each direction and by requiring that 
their velocity fits have a $\tilde{\chi}^2$ equal to one; this method produces positional uncertainties for IRS 21 that are
consistent with those found in Ghez et al. (1998), who explicitly measured the positional uncertainties. 
Table~\ref{pmsrcs} gives the proper motions and their uncertainties
as well as the positions of the sources at a specified epoch, while 
Figure~\ref{pmpl} shows the positions of the sources with respect to time. 
Compared to our initial report of IRS 21's proper motion (Ghez et al. 1998), the value given here is consistent, but has a factor
of five smaller uncertainty due to the additional epochs of data. For the other sources, these are
the first reported estimates of their proper motions. Table~\ref{pmsrcs} 
also provides the positions and proper motions of the Northern Arm 
sources estimated by Ott (2004).  We will use the uncertainty-weighted average of 
these two sets of observations in the subsequent calculations in this 
paper. 

\subsection{Model of the Northern Arm}

In order to explore the origin of the bowshocks as is done in $\S$5, we need a kinematic
model of the flow of the Northern Arm. Table~\ref{pmna} presents the values used, which come
from the Paumard et al. (2004) three-dimensional flow model which incorporates a series of 
non-planar Keplerian orbits. The velocities given in Table~\ref{pmna} represent the proper motion of the material at the 
position of each Northern Arm source as predicted by this model. The uncertainties provided are the
half range in each parameter which results in a 2$\sigma$ variation in the $\chi^2$ value.

\section{Discussion}

The observed extended emission demonstrates that the
extreme redness of these sources is, in part, caused by dust that is not
intrinsic to the source, but is rather from the postshock region of a 
bowshock, as suggested by Tanner et al. (2002). With the measured bowshock geometry and proper motions for some of the
Northern Arm sources, as well as the model-derived velocity vectors
of the Northern Arm material, it is now possible to investigate the origin of the bowshocks. 
A comparison of the measured proper motions of the sources (Table~\ref{pmsrcs}) 
with the flow of the Northern Arm (Table~\ref{pmna}), shows that they do not
agree. These observations, therefore, further rule out the hypothesis that the Northern Arm sources 
are recently formed stars that are being transported
inward in the Northern Arm flow, thus relieving the challenges presented by the short dynamical timescales
associated with the Northern Arm. The lack of any bowshock alignment toward the direction of Sgr A* allows us to rule out the formation of the
bowshocks as being due to the wind produced by the black hole. 
We, therefore, conclude that the observed bowshocks are generated
by an interaction with the Northern Arm flow, as opposed to any other medium.
We now consider a scenario in which
the embedded sources are interacting with this material and can calculate the
relative velocity vectors between the sources and the flow.  
In this case, the PAs of the resultant velocity vectors should correspond to the predicted PAs for bowshocks caused 
by an interaction with the Northern Arm. These
values are given for IRS 1W and 10W in Table~\ref{rel} and are consistent with the values based
on the morphology of the bowshocks given in Table~\ref{bowprop}. While the observed PA of IRS 10W is
marginally consistent with what is expected due to the interaction with the Northern Arm, it is
extended in the direction of the known red giant, 10EE, suggesting a possible interaction
with the winds from that star. 

Standoff distances measured for the Northern Arm sources provide some insight
into the spectral type of the underlying star.  The luminosity
of these sources (10$^{4-5}$ L$_{\sun}$, Tanner et al. 2001; Becklin et al. 1978) suggests that the most likely source types
for the central objects are 
AGB stars (v$_{w}$=10 km s$^{-1}$ and $\dot{m}$=10$^{-5}$ M$_{\sun}$/yr,  Lamers \& Cassinelli 1999), 
main sequence O stars (v$_{w}$=2000 km s$^{-1}$ and $\dot{m}$=10$^{-7}$ M$_{\sun}$/yr, Prinja, Barlow and Howarth 1990; Howarth and Prinja 1989) 
or WR stars (v$_{w}$=1000 km s$^{-1}$ and $\dot{m}$=2$\times$10$^{-5}$ M$_{\sun}$ yr$^{-1}$, Abbott \& Conti 1987). 
Table~\ref{rel} lists the standoff distances ($R_{AGB}$, $R_{O}$, and  $R_{WR}$ respectively) for all these 
stars based on the standoff distance relationship 
$R_o$ = 1.74$\times$10$^{19}$$\dot{m_*}$$^{1/2}$v$_w^{1/2}$v$_{rel*}^{-1}$$\mu_H^{1/2}$n$_H^{-1/2}$ cm (Weaver 
et al. 1977), 
where $R_o$ is the standoff distance, $\dot{m_*}$ is the stellar mass loss 
rate in units of 10$^{-6}$ M$_{\sun}$ yr$^{-1}$, v$_w$ is the 
stellar wind velocity in units of 10$^8$ cm s$^{-1}$, v$_{rel*}$ is the motion of the star relative to the
Northern Arm in units of 10$^6$ cm s$^{-1}$, $\mu_H$ is the mean molecular weight, and n$_H$ 
is the gas density of the Northern Arm in units of cm$^{-3}$, where we use a value of 3$\times$10$^4$ cm$^{-3}$ as
derived by Tanner et al. (2002).  
Only those standoff distances expected for the WR star scenario are consistent with the 
observed standoff distances. 
These large stand-off distances give the Northern Arm sources wind properties that are similar to 
the He I emission-line stars observed in and around the central parsec (Krabbe et al. 1995; Genzel et al. 2000; Paumard et al. 2001, 2003).

The distinct bowshock structure seen around IRS 8, combined with its 
projected distance from Sgr A*, has
led to the hypothesis that it has been flung out of the central parsec 
by a gravitational interaction with the supermassive black hole 
(Geballe et al. 2004). Since the diffuse emission associated with IRS 8, which has been  
associated with the ionized gas at the northern tip of the Northern Arm 
(Lacy et al. 1979), is infalling toward Sgr A*, 
the large observed standoff distance between the star and the gas and dust 
can be explained by the star's large projected distance from Sgr A*.
(40$\arcsec$). The large projected distance of IRS 8 from the supermassive
black hole means a smaller infall velocity for the surrounding diffuse
material. As a result there is a smaller relative velocity between the 
star and diffuse material producing a larger standoff distance around IRS 
8 than that seen around the embedded sources nearer to Sgr A*. This 
eliminates the requirement for the star itself to have a large proper motion, 
as is implied by the hypothesis that is has been flung out of the central 
parsec (Geballe et al. 2004).
It has also been suggested that the near-infrared emission from the 
bowshock around IRS 8 is due to
thermal shock emission from small grains (Geballe et al. 2004). However, 
we have attributed the near-infrared
emission around IRS 21 to scattered light, using a model of its SED 
(Tanner et al. 2002). Both IRS 21
and IRS 1W have large 2.2 $\micron$ polarizations which are larger than that expected from
the intervening ISM. They also have polarization position angles at 
2.2 $\micron$ which are deviant from the position angles 
of surrounding stars (Ott et al. 1999). 
Both scattered and thermally emitted light from grains aligned in the 
asymmetric outflows of these stars provide an explanation 
for the observed near-infrared polarization properties of these 
stars.

By modeling the asymmetric dust morphology of the Northern Arm sources 
as bowshocks from stars with wind velocities on the order of 1000 km s$^{-1}$, 
we can add IRS 1W, 10W, and 21, and most probably IRS 5 and 8, to the 
population of hot, windy stars within the
central parsec having winds and mass-loss rates resembling those of WR stars. 
Within a 24$\times$24 arcsecond region around the 
central parsec, a total of 15 stars have been confirmed as being 
candidate WR stars with strong winds (v$_{wind}$=300-1000 km s$^{-1}$) 
through their broad He I emission-line profiles (Paumard et al. 2003, 2001; 
Figer, private comm.).  With the width of the Northern Arm estimated from 
the FWHM of its emission at 12.5 $\micron$ (Tanner et al. 2003) and
the depth of the Northern Arm estimated from the half range of the 
line-of-sight distances of
IRS 1W, 5, 10W and 21 taken from the model of Paumard et al. (2004), we 
estimate a volume filling factor of the Northern Arm of 10\% within the 40$\times$24 arcsecond field of
view of the Paumard et al. (2001, 2004) He emission-line survey. 
Hence, given the known population of He I emission-line stars, we expect just a couple luminous stars to be coincident with the Northern Arm. 
Thus, the presence of three such stars (IRS 21, 1W, and 10W) and probably more (IRS 2 and 5) within this same region 
is slightly larger than expected given the observed population of He I emission-line 
stars suggesting the population of windy stars has been underestimated to date. 

The presence of OB and WR stars, which are believed to be massive and quite young, within a parsec of the central supermassive black hole is 
difficult to explain. Strong tidal forces and low present gas densities make {\it in situ} star formation unlikely, 
while formation at larger radii and subsequent inward migration is challenging given the short dynamical timescales. Nonetheless, clear signatures of tangential anisotropy in the He I emission-line stars suggest that these are indeed a very 
young, unrelaxed population. Furthermore, for the subset of those stars 
with three-dimensional velocity vectors and two-dimensional
position vectors, it has been shown that the orbits of the He I emission-line stars are consistent with at least
one and possibly two rotating planes (Levin \& Beloborodov 2003; Genzel et al. 2003). The identification of these
planes, however, is model dependent since the He I emission-line stars have only five of the six position-velocity
coordinates necessary to define the star's orbital planes.

The 
Northern Arm sources offer the first opportunity to test the validity of these planes since
all six phase-space values are known. Based on the proper motion data alone, IRS 1W and 10W are found to be moving in a 
clockwise direction around Sgr A*. We estimate the radial velocities of IRS 1W and 10W 
(see Table~\ref{pmsrcs}) from the derived inclination angle of the 
bowshock, the velocity of the gas, 
and the relative proper motion of the gas and the star. The line-of-sight distance from Sgr A* is
taken to be the same as the distance of the Northern Arm as given by the model from Paumard et al. (2003). 
Adding the measured proper motions and on-sky offsets from Sgr A*, we implement all six phase-space coordinates to
estimate the angles between the planes described by orbits of IRS 1W and 
10W (see Table~\ref{orbit}) and the planes of He I emission-line stars given 
in Genzel et al. (2003). Using the same line-of-sight normalized angular 
momentum as Genzel et al. (2003), J$_z/$J$_z$(max), we determine that both 
IRS 1W and 10W are on clockwise orbits. 
IRS 1W or 10W both lie in the same plane as the clockwise plane of
He I emission-line stars (see C angle in Table~\ref{orbit}), suggesting
that the Northern Arm sources, also being He I emission-line stars, share the same 
orbital plane as the known He I emission-line stars indicating that these 
disk hypotheses are valid for these sources.                                            

\section{Summary and Conclusions}

With high-resolution images revealing bowshocks around a population of luminous, embedded Northern Arm sources, we are able to 
conclude that they are all windy Wolf-Rayet stars and are thus likely to be members of the population of 
broad-line He I emission-line stars observed in the nearby IRS 16 cluster. 
They are not YSO's as previously speculated (Krabbe et al. 1995). 
The resolved bowshock structures are created from the interaction of the intrinsic stellar winds
and the flow of material along the Northern Arm. While IRS 1W and 10W show stellar wind properties
and kinematics similar to the known population of clockwise orbiting He emission-line stars, their
fully determined orbital parameters are coplanar with these sources. 
This supports the hypothesis that the He emission-line stars occupy 
at least one coherent orbital plane.  The Northern Arm sources offer 
a unique opportunity to test these plane models, so the addition of 
the known orbits of the remaining embedded sources (IRS 5, 8 and 2) 
would be beneficial. 

To further classify these Wolf-Rayet stars and attempt to directly measure their radial velocities,
deeper spectroscopic observations with high spatial resolution in the near-infrared would be worthwhile, with the goal of
finding the obscured spectroscopic signature of the stellar photosphere or wind. 
Chiar et al. (2003) observe the 6.2 $\micron$ PAH (circumstellar amorphous carbon) feature toward
the Quintuplet sources, which is known to be observed in intrinsic circumstellar dust shells around late type WR stars (DWCLs).  
If any absorption lines intrinsic to the star are detected, the resulting radial velocity 
measurements of the Northern Arm sources would also be a powerful tool
in understanding the dynamics of the He emission-line star population since their location within
the Northern Arm allows us to estimate their full phase space coordinates with no assumptions 
about their line-of-sight distance. There have been recent 3.8 $\micron$ observations of additional 
sources with bowshock morphologies within the central parsec, though these sources lack a 2.2 $\micron$ extended counterpart (Clenet et al. 2003).
We are just beginning to understand the significance of these sources and
with a more complete census of the region will be able to further investigate the population of windy stars
within the central parsec through their interaction with the dust and gas within the mini-spiral.

\begin{acknowledgements}
The work was supported through NSF grants No. AST-9988397 by the National Science Foundation Science 
and Technology Center for Adaptive Optics, managed by the University of
California at Santa Cruz under cooperative agreement No. AST-9876783 and the Packard Foundation.  
We thank Thibaut Paumard for
the use of his model of the flow of the gas along the Northern Arm and also Eric
Becklin for his many useful comments and ideas. 
Data presented herein were obtained at the W.M. Keck Observatory, 
which is operated as a scientific partnership among the California Institute of Technology, 
the University of California and the National Aeronautics and Space Administration. 
The Observatory was made possible by the generous financial support of the W.M. Keck Foundation.
The authors wish to recognize and acknowledge the very significant cultural role and 
reverence that the summit of Mauna Kea has always had within the 
indigenous Hawaiian community.  
We are most fortunate to have the opportunity to conduct observations from this mountain.
\end{acknowledgements}

\begin{appendix}
\section{IDAC and PSF Subtraction}

Since IDAC solves for both the object's spatial intensity distribution function 
as well as the PSF, it has more flexibility in the way it can be run than most deconvolution
algorithms. 
Prior to deconvolution, an initial estimate of the target image is obtained from 
the mean of the input images, and an
initial estimate of the PSF is created from a two-component Gaussian
fit to a reference star in the field. The latter is a reasonable
model of the PSF, modeling both the central diffraction-limited core and the broader, lower-intensity
residual seeing halo. The IDAC algorithm is initially run on the data cube of three to five 
images for $\sim$200 iterations, allowing
both the target and PSF images to vary. During this process, a point source in the field does
not reduce to a delta function in the target reconstruction image, requiring that the
reconstructed PSF be reconvolved with an isolated point source in
the target image to produce a new estimate for the PSF. 
The algorithm is then restarted for a couple hundred more iterations with the new PSF and 
target image from the previous deconvolution. This procedure is repeated a couple of times until a
good PSF is obtained for each frame within the data cube. The algorithm is then run for about
20 iterations with the PSF held fixed to remove any residual PSF structure
from the target estimate. Finally, the algorithm is run for a few more iterations with both the target
and PSF permitted to vary in order to relax the solution thus producing the final deconvolved image and PSF estimate.

IDAC has a known tendency to produce point sources whose size depends very modestly on their brightness.
Over five magnitudes, the core size is observed to increase by 35\% (Christou et al. 2004 in prep). This effect is negligible compared to the stand-off
distances observed in the Northern Arm sources, but needs to be accounted for in order to carry out PSF subtraction (\S 4.1). 
Since the Northern Arm sources are all brighter than their nearby point sources, the later are convolved with a gaussian to provide the best
match for PSF subtraction. In order to avoid contamination from the bowshock emission the comparison of core sizes is carried 
only out to source radii that correspond to 70\% of the peak intensity level. Figure~\ref{allbow} shows images of the Northern Arm sources
deconvolved with IDAC and PSF subtracted using the Gaussian convolved point source. 

\end{appendix}

\newpage

\begin{deluxetable}{lll}
\tablecaption{Summary of Observations\label{obs}}  
\tablehead{
\colhead{Telescope/Camera} & \colhead{Date} & \colhead{Sources}  
}
\startdata
Keck I/NIRC &  1995 June & 21$^a$ \\
Keck I/NIRC &  1996 June & 21$^a$ \\
Keck I/NIRC &  1997 May & 21$^a$ \\ 
Keck I/NIRC &  1998 May 14-15  &  1W,10W,21$^a$ \\
Keck I/NIRC &  1998 August 4-6 &  1W,10W,21$^a$ \\
Keck I/NIRC &  1999 May 2-4    &  1W,10W,21$^a$ \\
Keck I/NIRC &  1999 July 24-25 &  1W,10W,21$^a$  \\
Keck I/NIRC &  2000 May 19-20  &  1W,10W,21$^a$ \\
Keck I/NIRC &  2000 July 19-20  &  1W,10W,21$^a$ \\
Keck I/NIRC &  2001 May 9      &  1W,10W,21$^a$  \\
Keck I/NIRC &  2001 July 28-29 &  1W,10W,5,21$^a$ \\
Keck I/NIRC &  2002 April 23-24 & 21  \\ 
Keck I/NIRC &  2002 May 23-24 & 21\\
Keck I/NIRC &  2002 July 19-20  & 21 \\ 
Keck I/NIRC &  2003 April  21-23 & 21\\
Keck I/NIRC &  2003 July  22 & 21\\
Keck I/NIRC &  2003 September 7-8  & 21\\
Gemini/Hokupa'a-QUIRC & 2000 July 6 & 8  \\
\enddata
\tablenotetext{a}{2.2 $\micron$ observation of IRS 21 were presented originally by Tanner et al. 2002. Deconvolutions of these observations are introduced here along with proper motion measurements.}
\end{deluxetable}

\newpage

\begin{deluxetable}{lrrrr|rrr}
\footnotesize
\tablecaption{Azimuthal Dependence of Bowshock Morphologies Around the Northern Arm Sources\label{radii}}
\tablehead{
\colhead{Source} & \colhead{PSF Source$^a$} & \colhead{PA} & 
\colhead{Intensity} & \colhead{Radius [AU]} & \colhead{PA} &
\colhead{Intensity} & \colhead{Radius [AU]}
}
\startdata
IRS 1W &  1NE(3) &   10  & 0.530$\pm$0.005&  470$\pm$80 & 150 & 0.591$\pm$ 0.003&   710$\pm$140\\
&&      30  & 0.556$\pm$0.003&   460$\pm$90 & 250 &0.450$\pm$0.003&720$\pm$90\\
&&       50  & 0.520$\pm$0.003&   470$\pm$100 & 270  &0.762$\pm$0.003&620$\pm$80\\
&&       70 &  0.521$\pm$0.003&   470$\pm$100 & 290  &0.520$\pm$0.003&580$\pm$80\\
&&       90  &0.749$\pm$0.003&    490$\pm$90  & 310  &1.000$\pm$0.003&560$\pm$80\\
&&       110 &0.768$\pm$0.003&    460$\pm$100 & 330  &0.969$\pm$0.003&500$\pm$80\\
&&       130  &0.610$\pm$0.003&   570$\pm$120 & 350  &0.774$\pm$0.004&480$\pm$80\\
\hline
IRS 5  & -1.41+0.21 &  -10   & 0.856$\pm$ 0.004&  560$\pm$100 & 90 &0.654$\pm$ 0.004&510$\pm$90\\
&&       20 & 0.637$\pm$ 0.004&    470$\pm$100 &115&0.840$\pm$0.004&520$\pm$90\\
&&       40  & 0.351$\pm$ 0.004&    440$\pm$100 & 140 & 1.000$\pm$0.004&610$\pm$90\\
&&       70  &0.404$\pm$ 0.004&     460$\pm$90 &&& \\
\hline
IRS 8  & +1.52-5.0  & 35 & 0.437$\pm$ 0.004&   2030$\pm$200 & 270 &0.408$\pm$ 0.002&   2070$\pm$140\\
&&       40 & 0.629$\pm$ 0.003&   1980$\pm$180 & 280 & 0.492$\pm$0.003&   1940$\pm$160\\
&&       50 & 0.889$\pm$ 0.003&   1940$\pm$160 & 290 & 0.435$\pm$0.003&   1820$\pm$180\\
&&       60 & 1.000$\pm$ 0.003&   1920$\pm$160 & 300 & 0.426$\pm$0.003&   1700$\pm$160\\
&&       65 & 0.946$\pm$ 0.003&   1910$\pm$160 & 310 & 0.441$\pm$0.003&   1600$\pm$160\\
&&       75 & 0.818$\pm$ 0.003&   1910$\pm$160 & 315 & 0.428$\pm$0.003&   1520$\pm$160\\
&&       80 & 0.656$\pm$ 0.003&   1960$\pm$160 & 320 & 0.412$\pm$0.003&   1500$\pm$160\\

\hline
IRS 10W & 10EE  & 30 & 0.526$\pm$ 0.004&460$\pm$100 & 120 &1.000$\pm$0.004&410$\pm$100\\ 
        &       &  60 &0.675$\pm$ 0.004&400$\pm$80 & 150 &0.883$\pm$0.004&520$\pm$120\\
        &       &  90 &0.965$\pm$ 0.003&430$\pm$190 &180 &0.628$\pm$0.005&760$\pm$50 \\
\enddata
\tablenotetext{a}{PSF Sources for IRS 5 and 8 are labeled as the North-East offsets from the extended sources in arcseconds.}
\end{deluxetable}

\newpage

\begin{deluxetable}{lrrr}
\tablecaption{Measured Bowshock Properties\label{bowprop}}
\tablewidth{25pc}
\tablehead{
\colhead{Source} & \colhead{Standoff Distance [AU]} & \colhead {Inclination [\arcdeg]} & \colhead{PA [\arcdeg]} 
}
\startdata
IRS 1W$^a$ &  210$\pm$140   & 90$\pm$10 & 30$\pm$20  \\
IRS 1W &  460$\pm$160   & 90$\pm$10 & 30$\pm$30 \\
IRS 5  &  450$\pm$60  & 80/100$\pm$20 & 60$\pm$20 \\
IRS 8  & 1380$\pm$300 & 90$\pm$10 & 0$\pm$10   \\
IRS 10W & 420$\pm$60 & 70/110$\pm$20 & 85$\pm$20 \\
IRS 21$^a$ & 670$\pm$40  & \nodata  & \nodata \\
\enddata
\tablenotetext{a}{Values derived from the best fitting bowshock model after moving the apex 200 AU to the North}
\tablenotetext{b}{IRS 21 standoff distance as estimated by Tanner et al. (2002)}
\end{deluxetable}

\newpage

\begin{deluxetable}{lcrrrrrr}
\tablecaption{Kinematics of the Northern Arm Sources\label{pmsrcs}}  
\tablehead{   
\colhead{Source} & \colhead{t$_o$} & \colhead{r$_{\alpha t_o}^{a}$} & \colhead{r$_{\delta t_o}^{a}$} & \colhead{r$_z^b$}&\colhead{v$_\alpha^{a}$}  & \colhead{v$_\delta^{a}$} & \colhead{v$_z^c$} \\
\colhead{}       & \colhead{year}  &\colhead{arcseconds} & \colhead{arcseconds} & \colhead{arcseconds}& \colhead{km s$^{-1}$} & \colhead{km s$^{-1}$} &  \colhead{km s$^{-1}$}
}
\startdata
IRS 1W & 1999.164 & 5.283$\pm$0.007 & 0.571$\pm$0.008  & 7.537$\pm$1.536& -170$\pm$30 & 320$\pm$60 & 20$\pm$110 \\
IRS 10W & 1999.905 & 6.492$\pm$0.009 & 5.108$\pm$0.011 & 6.287$\pm$1.447 
&-160$\pm$110 & 10$\pm$210 & 175$\pm$130  \\
IRS 21 & 2000.207 & 2.383$\pm$0.006 & -2.713$\pm$0.007 & 4.455$\pm$1.080 & -300$\pm$30 & 100$\pm$30  & ... \\
\hline
IRS 1W$^d$ &         & 5.251$\pm$0.007 & 0.580$\pm$0.009 &   & -100$\pm$29& 216$\pm$37 & 20$\pm$90\\
IRS 10W$^d$ &        & 6.507$\pm$0.004 & 5.117$\pm$0.004 &   & 54$\pm$15 & 176$\pm$15 & 215$\pm$160\\
\enddata
\tablenotetext{a}{Position and velocity on the plane of the sky measured from 2.2 $\micron$ images. Additional systematic sources of error in the positions arise from uncertainties in the plate scale and image rotation.} 
\tablenotetext{b}{r$_z$ is estimated from the models of the Northern Arm with the assumption that these sources are embedded in the Northern Arm. In this convention, the line of sight axis increases away from the observer.}
\tablenotetext{c}{v$_z$ is estimated from the bowshock modeling described in \S~\ref{morph}. }
\tablenotetext{d}{Position and velocity on the plane of the sky from Ott 
(2004). We use the weighted average of these two measurements in 
subsequent calculations.} \end{deluxetable}

\newpage

\begin{deluxetable}{crrr}
\tablecaption{Properties of the Flow of the Northern Arm \label{pmna}}
\tablewidth{25pc}
\tablehead{ 
\colhead{Location} &\colhead{v$_\alpha$} & \colhead{v$_\delta$}   & \colhead{v$_z$}  \\
\colhead{}         &\colhead{km s$^{-1}$} & \colhead{km s$^{-1}$} &  \colhead{km s$^{-1}$}
}
\startdata 
IRS 1W    & -70$\pm$50    &  -210$\pm$70 &   20$\pm$50\\
IRS 5     &  10$\pm$130   &  -140$\pm$190 &  110$\pm$60\\
IRS 10W   & -20$\pm$70    &  -190$\pm$100 &  80$\pm$60\\
IRS 21    & -130$\pm$60   &  -320$\pm$50  & -90$\pm$20 \\
\enddata
\end{deluxetable}

\newpage

\begin{deluxetable}{lrrrr}
\tablecaption{Predicted PAs and Standoff Distances for the Embedded Sources\label{rel}} 
\tablewidth{25pc}
\tablehead{   
 \colhead{Source} &  \colhead{PA$_{pred}^a$}& \colhead{$R_{AGB}^b$} & \colhead{$R_{O}^b$} & 
\colhead{$R_{WR}^b$}  \\
 \colhead{}       &  \colhead{degrees}      & \colhead{AU}          & \colhead{AU}      &   \colhead{AU}                        
}
\startdata
IRS 1W   & -10$\pm$10 &40$\pm$10 &60$\pm$10 & 560$\pm$40   \\
IRS 10W  &  10$\pm$10 &50$\pm$10 &70$\pm$10 &  680$\pm$30  \\
IRS 21   &  -20$\pm$10 & 40$\pm$10 &55$\pm$10& 550$\pm$30    \\
\enddata
\tablenotetext{a}{This is the measured PA of the relative velocity between the sources and the Northern Arm flow on the
plane of the sky and is expected to correspond to the PA of the bowshock if the bowshock is created by an interaction of the star's wind
and the Northern Arm material.}
\tablenotetext{b}{The uncertainties only incorporate our measured uncertainties for the relative velocities and stellar proper motions.}
\end{deluxetable}

\newpage

\begin{deluxetable}{cccccccc}
\footnotesize
\tablecaption{Estimated Orbital Planes \label{orbit}}
\tablehead{ 
\colhead{Source}   &\colhead{Observed i}&\colhead{Orbit i}  & 
\colhead{Orbit $\Omega$} & \colhead{J$_z/J_z(max)$} & \colhead{C angle$^a$} & \colhead{CC Angle$^a$}&\colhead{1W/10W angle}\\
\colhead{}         &\colhead{degrees}&\colhead{degrees} & \colhead{degrees} & \colhead{} & \colhead{degrees} & \colhead{degrees} & \colhead{degrees} 
}
\startdata 
IRS 1W    &90& 123$\pm$6    &  149$\pm$14 & 0.93$\pm$0.16 & 27$\pm$35 & 162$\pm$60& 47$\pm$53\\
IRS 10W   &70& 129$\pm$27   &  90$\pm$44 & 0.61$\pm$0.10  & 25$\pm$114 & 131$\pm$39 &  \\
IRS 10W   &110&121$\pm$15   &-155$\pm$38 &                & 73$\pm$31 & 139$\pm$43 & 47$\pm$43\\
\enddata
\tablenotetext{a}{C and CC stand for clockwise and counter clockwise.}
\end{deluxetable}

\newpage

\figcaption[f1.ps]{Contour 2.2 $\micron$ images of the 0$\farcs$5$\times$0$\farcs$5 region around 
the Northern Arm sources. Column 1 shows the raw images, Column 2 is a
corresponding PSF taken from the same image, Column 3 shows the images deconvolved using the IDAC deconvolution algorithm (Jefferies \& Christou 1992).
The contours plotted represent 90-20\% of the peak value in 10\% intervals. North is up and East is to the left.\label{nearall}}

\figcaption[f2.ps]{a) Contour image of IRS 5 along with lines depicting the arcs over which radial profiles were averaged. North is up and East is to the left.   b) 
Plots of those radial profiles for which the peak is within 70\% of the maximum value of all the profiles. \label{morphcalc}}

\figcaption[f3.ps]{Contour images of the two best fits of the bowshock model to the
radii extracted from the image of IRS 1W. Left: The best fit derived after shifting the
apex of the bowshock 200 AU to the North. Right: The best fit derived without shifting the
apex.\label{irs1wcomp}}

\figcaption[f4.ps]{Center: Greyscale image of the central 20$\times$20 arcseconds of the
galactic center at 12.5 $\micron$ presented in Tanner et al. (2002) overplotted with the symbols representing
the position of the Northern Arm stars (+), the broad-line He emission-line stars (diamonds) and
narrow-line He emission-line stars (*) from Paumard et al. (2001, 2003). 
Sides: Contour images of IRS 1W, 5, 8 and 10W from the deconvolved Keck/NIRC and Gemini/AO 2.2 $\micron$ 
data after PSF subtraction. Overplotted as crosses in the contour images are the positions of the peaks in the radial profile used to find the best fitting bowshock models 
(superimposed curve). North is up and East is to the left.  \label{allbow}}

\figcaption[f5.ps]{ Plots of the $\alpha$ and $\delta$ position of IRS 21, 1W, and 10W, 
as a function of time, along with the least squares fit to the data. \label{pmpl}}

\clearpage

\begin{figure}[ht]
\epsscale{0.75}
\plotone{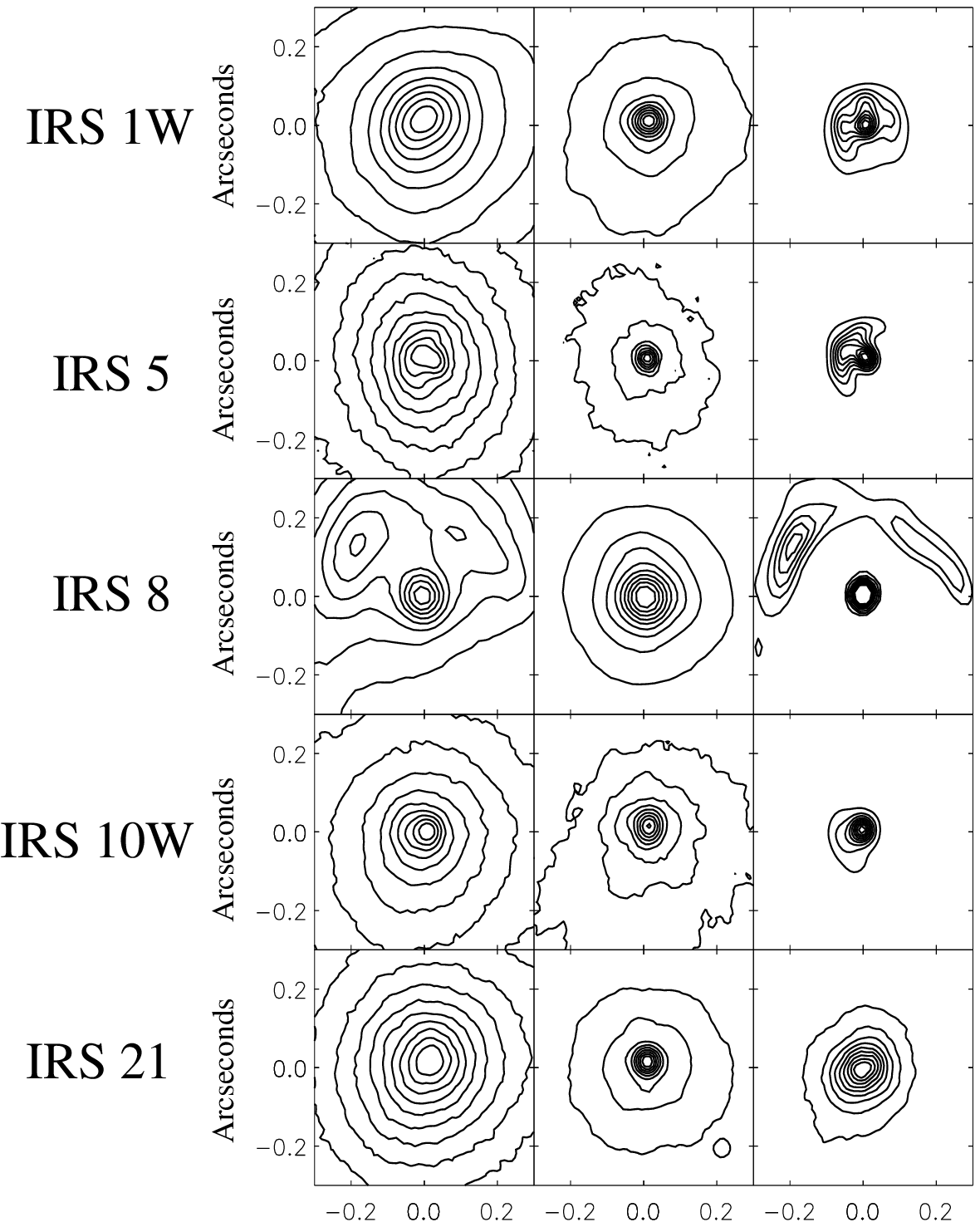}
\end{figure} 
\clearpage
\begin{figure}[ht]
\epsscale{1.0}
\plotone{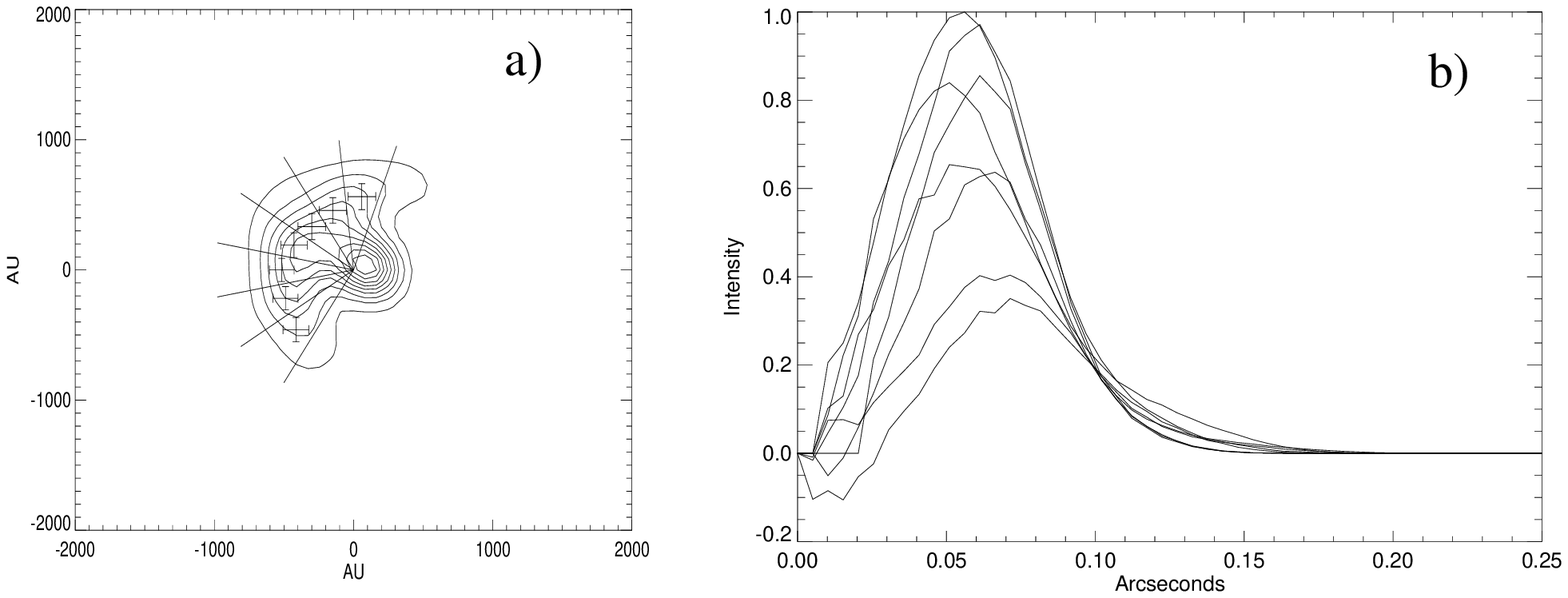}
\end{figure} 
\clearpage
\begin{figure}[ht]
\epsscale{1.0}
\plotone{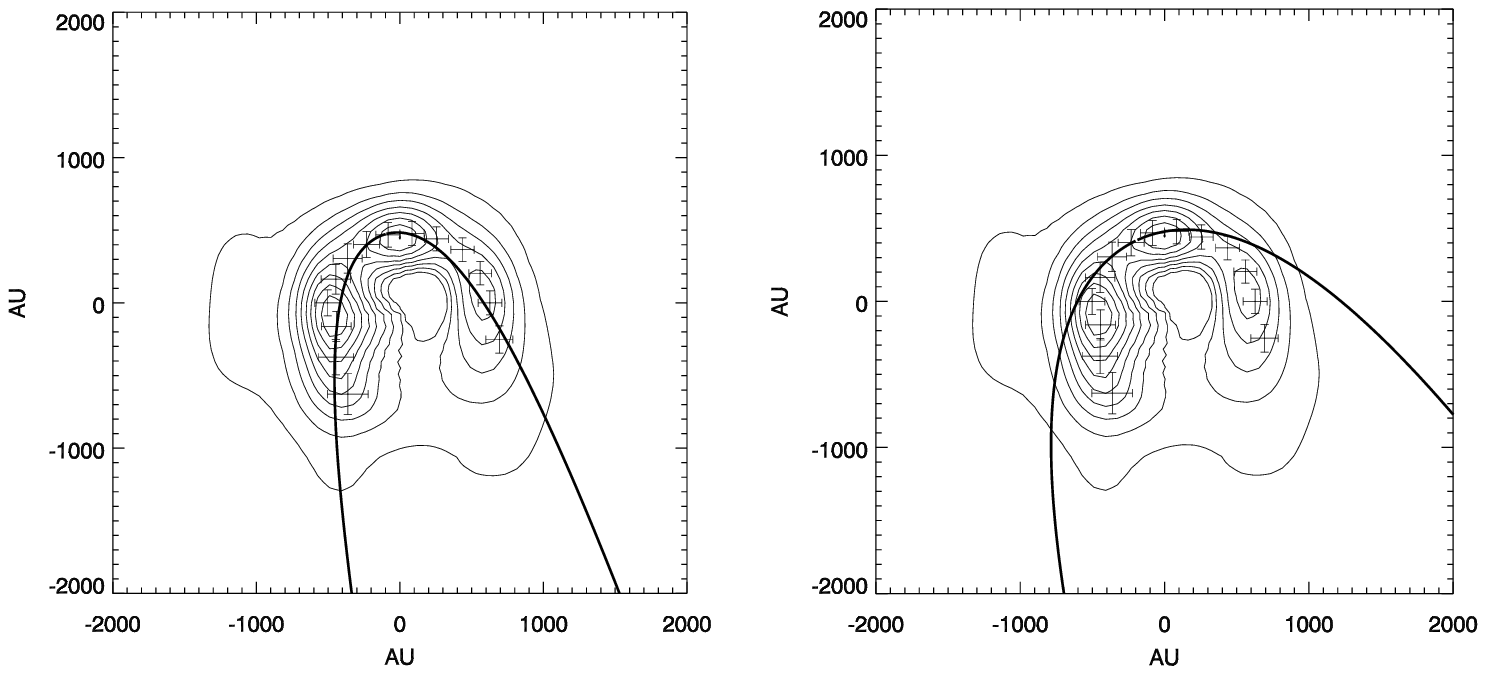}
\end{figure} 
\clearpage
\begin{figure}[ht]
\epsscale{1.0}
\plotone{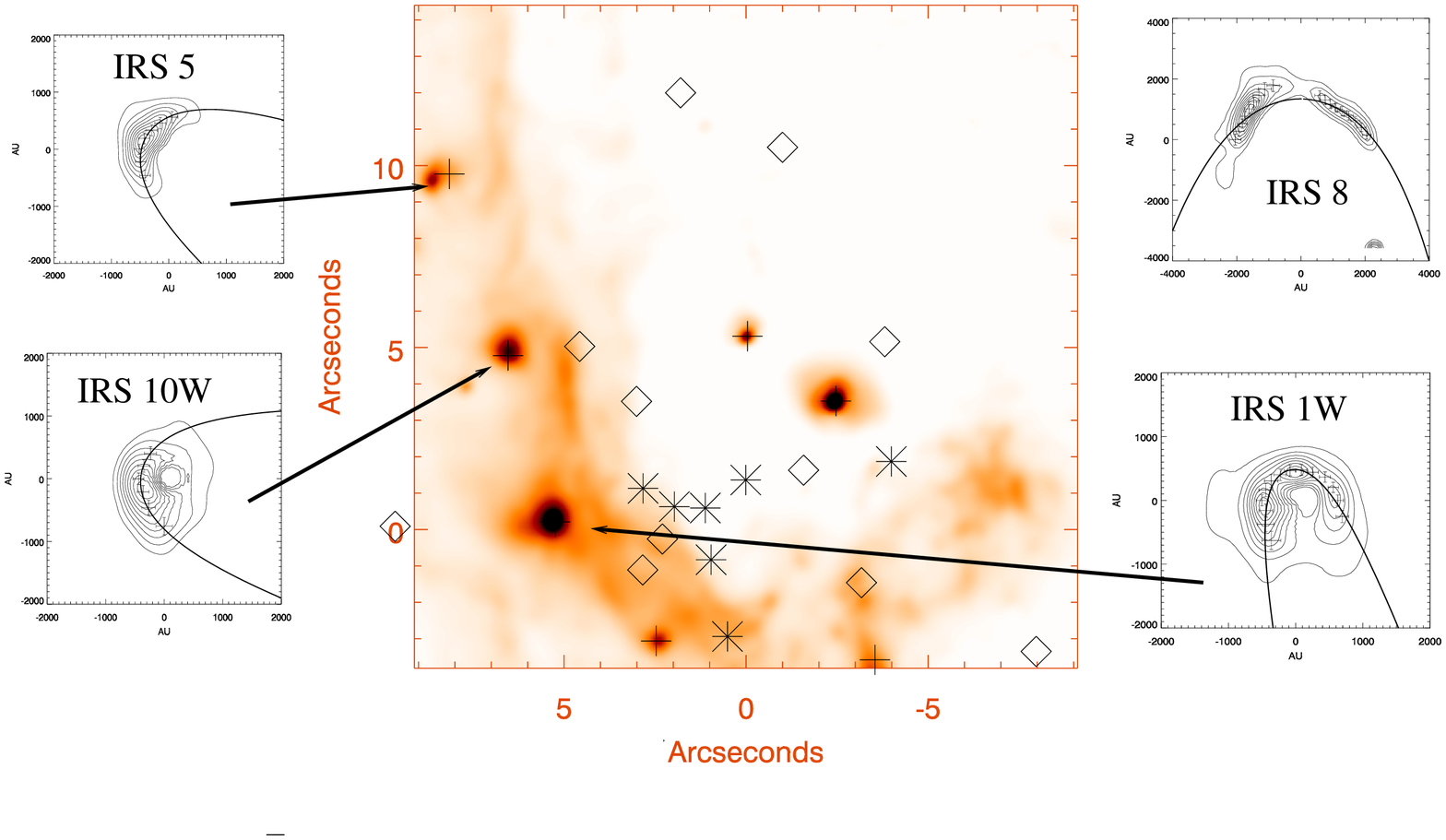}
\end{figure} 
\clearpage
\begin{figure}[ht]
\epsscale{1.0}
\plotone{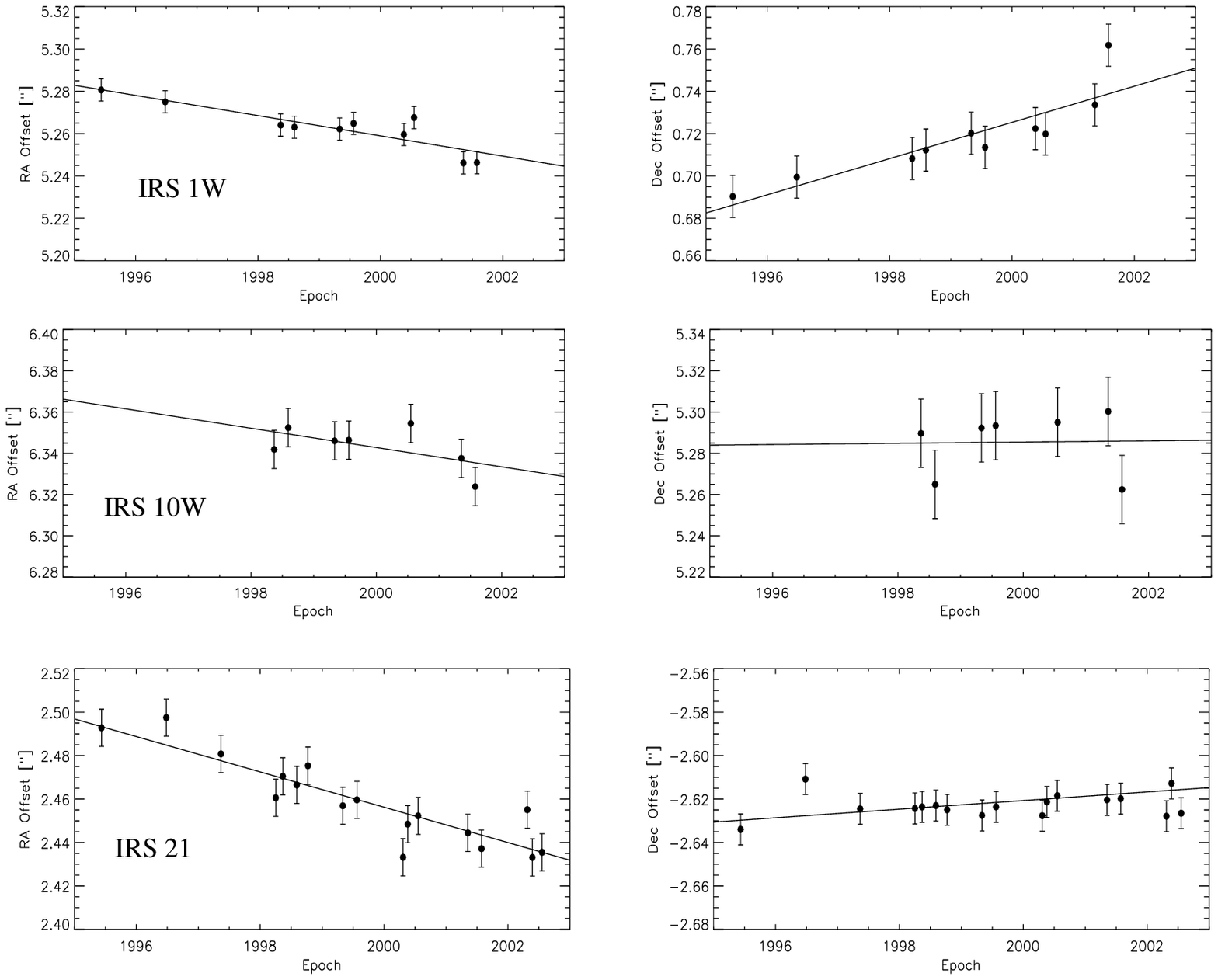}
\end{figure} 

\end{document}